\documentclass[12pt,letterpaper]{article}

\usepackage{amsmath,amssymb,calc}
\usepackage{graphicx}
\usepackage{color}

\renewcommand{\theequation}{\thesection.\arabic{equation}}

\newcommand\be{\begin{equation}}
\newcommand\ee{\end{equation}}
\newcommand{\bea}{\begin{eqnarray}}
\newcommand{\eea}{\end{eqnarray}}

\newcommand{\nn}{\nonumber}

\def\id{\protect{{1 \kern-.28em {\rm l}}}}

\def\1{^{(1)}}
\def\0{^{(0)}}
\def\2{^{(2)}}

\def\id{\protect{{1 \kern-.28em {\rm l}}}}

\numberwithin{equation}{section}
\begin{document}
\begin{titlepage}

\begin{center}
\hfill \\
\vskip 1cm 
{\Large \bf The Lifetime of Axion Stars}
\vskip 1cm 
{\it Joshua Eby, Peter Suranyi, and L.C.R. Wijewardhana \newline
Dept. of Physics, University of Cincinnati, Cincinnati, OH 45221, USA}
\end{center}
\vskip 2 cm
\begin{abstract}
 We investigate the decay of condensates of scalars in a field theory defined by $V({\cal A})=m{}^2\,f{}^2\,[1-\cos({\cal A}\,/\,f)]$, where $m$ and $f$ are the mass and decay constant of the scalar field.  An example of such a theory is that of the axion, in which case the condensates are called axion stars.  The axion field, $\cal A$,  is self adjoint. As a result the axion number is not an absolutely conserved quantity. Therefore, axion stars are not stable and have finite lifetimes. Bound axions, localized on the volume of the star, have a coordinate uncertainty $\delta x \sim R \sim 1/(m_a \Delta)$, where $R$ is the radius of the star and $\Delta = \sqrt{1-E_0{}^2\,/\,m_a^2}$.  Here $m_a$ and $E_0$ are the mass, and the ground state energy of the bound axion. Then the momentum distribution of axions has a width of $\delta p\sim m_a\,\Delta$. At strong binding, $\Delta={\cal O}(1)$, bound axions can easily transfer a sufficient amount of momentum to create and emit a free axion, leading to fast decay of the star with a transition rate $\Gamma \sim m_a$.  However, when $\Delta\ll 1$, the momentum distribution is more restricted, and as shown in this paper, the transition rate for creating a free axion decreases as 
 $\exp (-p \, \delta x) \sim \exp (-\Delta^{-1})$. Then sufficiently large, weakly bound axion stars, produced after the big bang, survive until the present time. We plot the region of their stability, limited by decay through axion loss and by gravitational instability, as a function of the mass of the axion and the mass of the star.
\end{abstract}
\end{titlepage}
\section{Introduction}
Axions \cite{firsta,firstb,firstc,firstd,seconda,secondb,secondc,secondd}, hypothetical particles proposed to solve the strong $CP$ problem, have been considered to be  candidates for dark matter  \cite{third,third-a,third-b,third-c,fourth}. Their cosmological history is highly nontrivial; various aspects, including production mechanisms and topological defects, have been analyzed previously \cite{Khlopov1,Khlopov2,Khlopov3,Khlopov4,Khlopov5}. Of particular interest is the idea that axions can condense in the early universe into axion stars through gravitation and a well defined self-interaction \cite{Tkachev,Kolb}.  

Ruffini and Bonazzola~\cite{RB} investigated boson stars in a free field theory of gravitating bosons by using equations of motion obtained from the expectation value of the Einstein equations and of the Klein-Gordon equation.  Following the method outlined in their paper, several authors (including us) ~\cite{Barranco,ESVW} have analyzed the structure of gravitationally bound axion stars in an axion field theory defined by the potential
\be
V({\cal A})= m_a{}^2\,f_a{}^2\left[1-\cos\left(\frac{{\cal A}}{f_a}\right)\right],
\ee
where $\cal A$ is the axion field, $f_a$ is the decay constant and $m_a$ is the mass of the axion. 
In reference~\cite{ESVW}  the equations of motion were solved using a systematic expansion in the binding energy and the ratio of $f_a\,/\,M_P$, where $M_P$ is the Planck scale. The weak binding mass spectrum of axion stars reaches a maximum mass at $\kappa = f_a{}^2 /M_P{}^2  \Delta^2  \approx 0.34$. In other words, no stable axion star solution exists with $M \gtrsim 25.5\,f_a\,M_P / m_a$, which corresponds to a maximum number of condensed axions $N\approx 25.5\,f_a\,M_P / m_a{}^2$. For QCD axion stars, this maximum mass corresponds to a Schwarzschild radius of $\mathcal{O}(10^{-8}$ m$)$, while the actual radii of such stars are $\mathcal{O}(100-1000$ km$)$; thus, they are very far from being or becoming black holes. Masses and radii of similar order were found by the authors of~\cite{Guth}, using a different method of analysis. If we try to increase their mass, e.g. by shooting two stable axion stars together, we suspect that the resulting unstable configuration would shed excess axions through short-distance repulsive interactions in the axion potential.

A problem, calling the existence of axion stars into question, is that the axion field is self-adjoint and axion field theory does not have a conserved  axion number. In view of the nontrivial self-interaction term in the action, it may be thought that primordial axion stars created in the early universe cannot survive to the present time.  In this paper we calculate the decay rate of axion stars, as a function of the parameters of the axion field theory coupled to classical gravity.  

Axions decay in a process $a\to 2\,\gamma$. This decay rate into photons was calculated and was found to be slow enough not to affect significantly the stability of axions or axion stars on a cosmological time scale~\cite{Tkachev-1}. 
However, due to self-interaction terms in the Hamiltonian, the reaction $3\,a\to a$ can also take place inside a condensed axion star, because the rest of the star can absorb arbitrary momentum from the three initial state axions. Axions bound inside the star are in the ground state with a  wave function that is clearly {\it not in a momentum eigenstate.}  In this paper, we will calculate the decay rate of axion stars due to this latter process. The rate depends on the self-coupling parameter $(m_a/f_a)^2$, and also on an additional  integration constant of the equations of motion. This can be chosen as the binding energy of axions, or any of the following quantities: the total mass of the star, the central density, or the number of axions in the star. 
 
 Recent papers investigating  axion stars~\cite{Guth,Braaten} introduce effective complex axion fields in the nonrelativistic limit, in which  matrix elements which are non-diagonal in axion number vanish, and as a result the axion number is conserved.  However, we know that because axions are real bosons, the axion number is not conserved, and consequently axion stars cannot live forever. To obtain an estimate of the decay rate of axion stars, one must go beyond effective theories that conserve  axion number.  In this paper we intend to propose a simple, and in our view natural, addition to axion number-conserving terms of the theory, that allows one to estimate the lifetime of axion stars.
\section{Equations of motion}
Ruffini and Bonazzola \cite{RB} used a quantized free Hermitian boson field for studying boson stars.  They postulated that the spherically symmetric gravitational field of the boson star has the metric
\be
ds^2=-B(r)\,dt^2 +A(r) \, dr^2+r^2\,d\Omega.
\ee

Omitting all but the contribution of the ground state, the   boson field takes the form
\be\label{field1}
\Psi= R(r)\,e^{i\,E_0\,t}\,a_0+\text h.c.
\ee
where $[a_0,a^\dagger_0]=1$, $E_0$ is the ground state energy, and $R(r)$ is a real-valued function. 
Ruffini and Bonazzola took the expectation value of the $rr$ and $tt$ components of the Einstein equations and of the  Klein-Gordon equation, between $N$-boson (and $N-1$ boson) states \cite{RB}. They numerically solved the resulting set of equations for $R(r)$, $A(r)$, and $B(r)$ to calculate the mass and radius of boson stars.  
  
  Barranco et. al.~\cite{Barranco} applied the method of ~\cite{RB} to axion stars.  Axions are described by the Klein-Gordon equation in a gravitational field with classical potential
\be\label{pot}
V({\cal A})=m_a{}^2\,f_a{}^2 \left[1- \cos\left(\frac{{\cal A}}{f_a}\right)\right]= \frac{1}{2}m_a{}^2\, {\cal A}^2-\frac{1}{24} \frac{m_a{}^2}{f_a{}^2}\,{\cal A}^4+...
\ee
Having  non-zero interaction terms complicates matters, as expectation values of the composite operators, appearing in the equations of motion, also have quantum corrections.  However, diagrams with additional vertices, which are attached to at least one internal line, are negligible,\footnote{ One can show that every internal line generates an additional factor of $m_a{}^2\,/\,f_a{}^2\ll 1$ in  matrix elements.} and solving the equations of motion generates appropriate tree level corrections. 

 In a recent paper~\cite{ESVW}, we investigated axion stars in the weak binding regime  using a quantum axion field, similar to (\ref{field1}).  We have shown that the expectation value of the  potential (\ref{pot}) in the ground state of $N$ particles gets modified to the tree level quantum potential
 \be\label{potential}
V(\Phi)=2\,m_a{}^2\,f_a{}^2\left[1-J_0\left(\frac{\Phi}{f_a{}}\right)\right]=\frac{1}{2}\,m_a{}^2\,\Phi^2-\frac{1}{32}\frac{m_a{}^2}{f_a{}^2}\Phi^4+...
\ee
where $\Phi=2\,\sqrt{N}\,R(r)$ \cite{ESVW}.  Here and in what follows, $J_k$ is the $k$th Bessel function of the first kind.
The weak binding limit is defined by $\Delta\ll1$ with
\be
\Delta=\sqrt{1-\frac{E_0{}^2}{m_a{}^2}}\simeq\sqrt{-2\,\frac{\delta E}{m_a}},
\ee
where $\delta E=E_0-m_a<0$ is the binding energy.
In this limit, solutions and physical quantities satisfy scaling relations such that, aside from overall multipliers of powers of $M_P$, $f_a$, $m_a$ and $\Delta$, they depend only on the dimensionless combination\footnote{The parameter $\lambda$ in \cite{ESVW} is related to $\kappa$ by $\lambda^2=\kappa$.}
 \be
 \kappa=\frac{f_a{}^2}{{M_P{}^2\,\Delta^2}}.
 \ee
 As an example for the power of the weak binding approximation, at $\Delta\ll1$ the boundary of the gravitational instability is at $\kappa\simeq 0.34$, irrespective of the value of other parameters~\cite{ESVW}.
 We will  use $\Delta$  to parametrize solutions at given $m_a$ and $f_a$. All other quantitites that describe the solution, such as the mass of the star ($M$), the number of condensed axions ($N$), and the central density, can be easily expressed as a function of $\Delta$.  It turns out that the regions of stability and many other physical properties have their simplest analytic expressions in terms of $\Delta$. In this paper, we  restrict ourselves to weak binding, as we find the transition to instability sets in at fairly small values of $\Delta$.

The system of equations satisfied by the three dimensionless fields, $A,\,B,$ and $Z$ are \cite{Colpi,RB,Barranco}
\bea\label{eom0}
\frac{A'(y)}{A(y)}&=&\frac{1-A(y)}{y}+\frac{y\,A(y)}{4\,\Lambda}\left\{\frac{\epsilon_0{}^2\,Z(y)^2}{B(y)}+\frac{Z'(y)^2}{A(y)}+4[1-J_0(Z)]\right\},\nn\\
\frac{B'(y)}{B(y)}&=&\frac{A(y)-1}{y}+\frac{y\,A(y)}{4\,\Lambda}\left\{\frac{\epsilon_0{}^2 \, Z(y)^2}{B(y)}+\frac{Z'(y)^2}{A(y)}-4[1-J_0(Z)]\right\},\nn\\
Z''(y)&=&-\left[\frac{2}{y}+\frac{B'(y)}{2\,B(y)}-\frac{A'(y)}{2\,A(y)}\right]Z'(y)-A(y)\left[\frac{\epsilon_0{}^2\,Z(y)}{B(y)}-2\,J_1(Z)\right],
\eea
where we introduce a dimensionless coordinate $y=m_a\,r$, rescaled energy $\epsilon_0 = E_0\,/\,m_a$, and dimensionless field $Z(y)=\Phi(r)\,/\,f_a=2\, \sqrt{N}\,R(r)\,/\,f_a$.   Furthermore, we define the parameter\footnote{The parameter $\delta$ in \cite{ESVW} is related to $\Lambda$ by $\delta = \Lambda^{-1}$} $\Lambda=M_P{}^2\,/\,f_a{}^2$.

The approximation $\Delta\ll1$ defines the weak binding limit~\cite{ESVW}.  In this limit, the gravitational field is also weak, so we can expand the metric components as $A(x)\simeq1+ a(x)\,/\,\Lambda$ and $B(x)\simeq 1+b(x)\,/\,\Lambda$, where the coordinate is  $x=\Delta\,y$. Furthermore, the axion field rescales as $Z(y)=\Delta\, Y(x)$.

Then in leading order of $\Delta$ and finite $\kappa$ the equations of motion reduce to 
\bea\label{eomY}
Y''(x)&=&[1+\kappa \,b(x)]Y(x)-\frac{2}{x}\,Y'(x)-\frac{1}{8}\,Y(x)^3,\nn\\
a'(x)&=&\frac{x}{2}\,Y(x)^2-\frac{1}{x}\,a(x),\nn\\
b'(x)&=&\frac{1}{x}\,a(x).
\eea
The physical significance of the constant $\kappa$ is clearly shown by (\ref{eomY}): it  is the effective coupling constant between the gravitational and axion fields.  Rather than $\Delta$ and $\Lambda$ separately,  the equations of (\ref{eomY}) depend only on $\kappa$. As we have shown in ~\cite{ESVW}, axion stars at $\kappa \simeq 0.34$ have maximal mass at fixed $m_a$, while stars with larger masses are gravitationally unstable. Further, stars at $\kappa \gtrsim 0.34$ are susceptible to tunneling to more strongly bound states of equal $N$ but lower total energy, at $\kappa \lesssim .34$.

A further simplification, the complete decoupling of gravity from the axion field, happens if in addition to $\Delta\ll1$, we also have $\kappa\ll1$. This occurs at sufficiently small decay constant of the axion, when the following inequalities are satisfied: $1\,/\,\Lambda\ll\Delta^2\ll 1$. Then the nonlinear Klein-Gordon equation for the axion field takes the form
\be\label{eomY0}
Y''(x)=Y(x)-\frac{2}{x}\,Y'(x)-\frac{1}{8}\,Y(x)^3.
\ee
The investigation of the singularity structure of the real analytic solution of (\ref{eomY0}), given in Appendix II, facilitates the calculation of decay rates of axion stars.

\section{Axion star decay}

\begin{figure}[hbt]
\centering \includegraphics{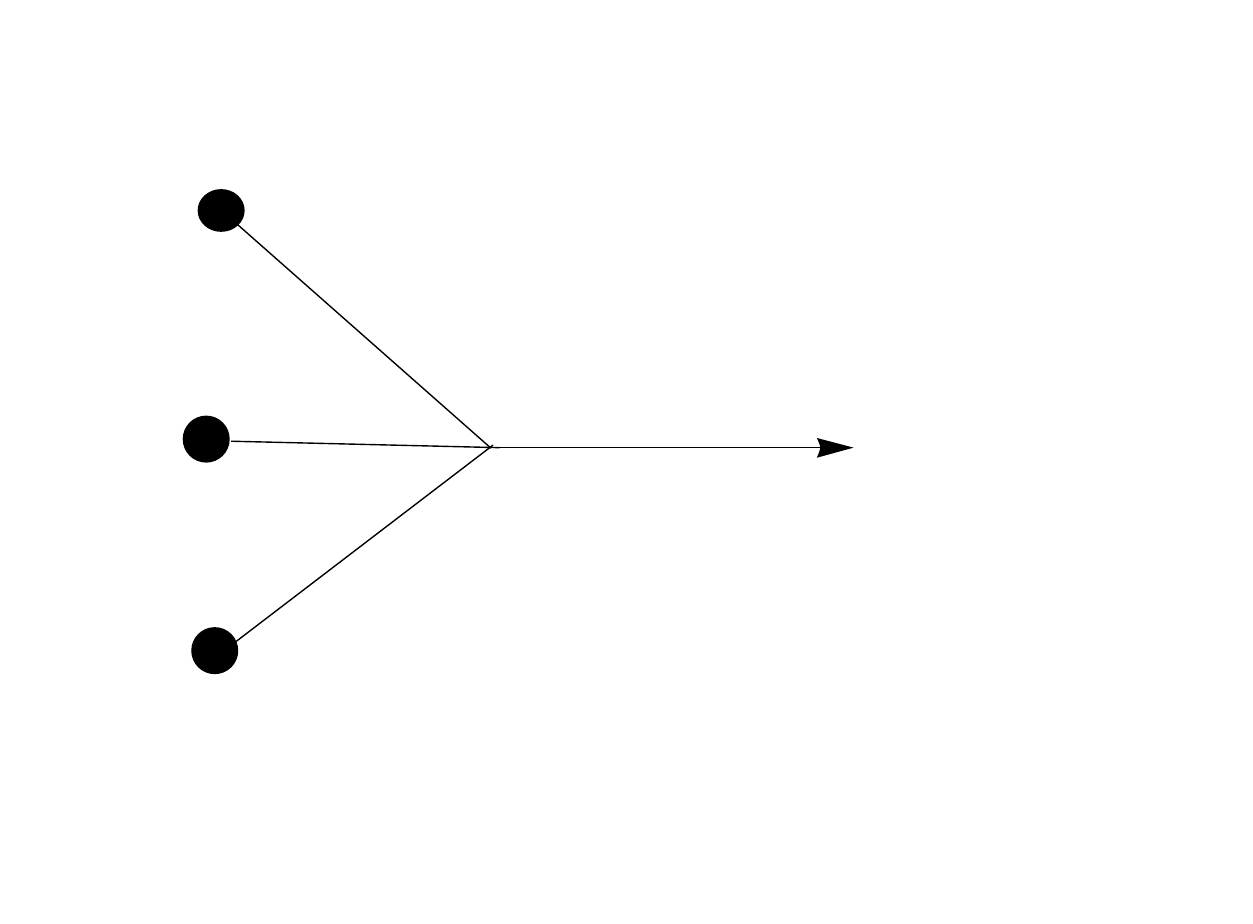}
\caption{A leading order diagram contributing to the decay of axion stars.   Lines anchored in a blob represent bound axions in the ground state of the axion  star, with a smeared momentum distribution, while the arrow represents a free axion with definite momentum.}\label{decaydiagram}
\end{figure}

Since the axion field is self-adjoint, there is no conservation law protecting the axion number.  The self-interaction terms of the axion potential lead to  the decay of axion stars. We will concentrate on the leading  contribution to the transition generated by the potential operator $V(\Phi)$ in (\ref{potential}). 
Among others, the axion potential contributes to the decay of axion stars via the process $3\,a\to a$.  The diagram for this process is shown in Fig. \ref{decaydiagram}. First order processes, like $(2\,n+1)\,a\to a$, where $n=2,\,3,...$, also contribute to the decay of axion stars, though we will show that their contribution is suppressed by an additional factor of $\Delta$ for each additional ground state axion absorbed.

It must be emphasized that  Fig. \ref{decaydiagram} {\it is not a standard Feynman diagram} in which all external particles are in definite momentum states. (They are, however in energy eigenstates.)  The "incoming" particles, which are bound axions, are localized in "wave packets" which are extended over the whole axion star, which has radius $R$. Therefore, their momentum distribution is spread over a range of $\delta p\sim 1\,/R$.  The momentum of the bound axions can be transferred to the emitted free axion.  In other words, the process depicted in Fig. \ref{decaydiagram} conserves energy but it does not conserve momentum. The excess momentum is transferred from the axion star, which has practically infinite mass, to the emitted particle. The remaining axion star suffers an infinitesimal recoil as a result.

It is thus possible for three bound axions to convert into a free axion, but only if they can transfer a momentum of $p=\sqrt{9\,E_0{}^2-m_a{}^2}=m_a\sqrt{8-9\Delta^2}$ to the rest of the condensate.  This is possible because the coordinate uncertainty of bound axions is $\delta x \sim 1/(m_a\,\Delta)$, and so their momentum distribution is spread over a range of $\delta p \sim m_a\,\Delta$.  At strong binding $\Delta=\mathcal{O}(1)$, it is easy to transfer sufficient momentum to the escaping axion.   As a result, the transition rate is  large and strongly bound axion stars decay soon after they are produced.  However, when the binding energy is small $\Delta\ll1$, it becomes more and more difficult to transfer sufficient momentum,  $p\simeq\sqrt{8}\,m_a$, to the created free axion.  We will see in Sec. \ref{lifetime_Sec} that the transition rate drops exponentially as $\Delta\rightarrow0$.  At a particular value of $\Delta$, the transition rate becomes so small that the axion star survives from the big bang until the present time.  It turns out that the transition from unstable to stable axion star occurs at $\Delta\simeq0.05-0.06$. As a result, while searching for the transition point, it is sufficient to  consider small binding energies.

To quantify the calculation of the transition rate, we propose  to modify the axion field (\ref{field1}) to include  a free axion term
\be\label{field2}
\Phi= R(r)\,e^{i\,E_0\,t}\,a_0+\int \frac{d^3p}{\sqrt{2\,E_p}} \,a_p \,e^{i\,p\cdot r-i\,E_p\,t}+\text h.c.
\ee
where $[a_p,a^\dagger_{p'}]=\delta^3(p-p')$.  Note that the two terms in (\ref{field2}) are not strictly orthogonal. However, the ground state wavefunction is sharply peaked at zero momentum (with width $\sim m_a\Delta$), and as we have pointed out, the momentum of the emitted axion in the leading-order process is sharply peaked at a very large momentum $p=\sqrt{8}m_a$. Thus, the overlap of the states represented by the two terms in (\ref{field2}) is considered to be negligible in what follows.

In leading order the transition rate is obtained from the matrix element of the interaction potential
\be\label{matrix}
{\cal M}_3=\int dt\,d^3r\,\left\langle N\left|V(\Phi)\right|N-3,p\right\rangle=m_a{}^2\,f_a{}^2\int dt\,d^3r\,\left\langle N\left|1-\cos\left(\frac{\Phi}{f_a}\right)\right|N-3,p\right\rangle,
\ee
where the operator $\Phi$ is given in (\ref{field2}) and where the states $\langle N\vert$ and $\vert N-3,p\rangle$ refer to the initial condensate composed of $N$ bound axions, and the final state composed of an axion star of $N-3$ bound axions plus a free axion of momentum $p$, respectively.
Using (\ref{field2})  and  $N\gg1$ we obtain the only non-vanishing contribution to (\ref{matrix}) as\footnote{More details about the derivation of (\ref{matrix}) and its generalization are given in Appendix I. The calculation of the expectation value of similar operators, between states $\langle N|$ and $|N\rangle$, or $\langle N|$ and $|N-1\rangle$, were derived in the appendix of~\cite{ESVW}.}
\bea\label{matrix2}
{\cal M}_3&=&-i\,\frac{m_a{}^2\,f_a}{\sqrt{2\,E_p}}\,\int dt\,d^3r\, J_3\left(\frac{2\,\sqrt{N}\,R(r)}{f_a}\right)\,e^{i\,p\cdot r}\,e^{i(3\,E_0-E_p)\,t}\nn\\
&=&-4\,i\,\pi^2\,\frac{1}{\sqrt{2\,E_p}}\delta(3\,E_0-E_p)\,\frac{f_a}{m_a\,k}\,I_3(k),
\eea
where
\be\label{Idef}
I_3(k)=\int_{-\infty}^\infty dy\,y\,\sin(k\,y)\,J_3[Z(y)]
\ee
and $k=p\,/\,m_a$.  We extended the range of integration in (\ref{Idef}) to $-\infty<y<\infty$ using the fact that 
$Z(-y)=Z(y)$.

 As we indicated in the previous section, at small binding energy $\Delta\ll1$, it is advantageous to use the scaling of \cite{ESVW}, namely $Z(y)=\Delta \,Y(x)$ and $x=\Delta\,y$, resulting in the following expression for $I_3(k)$:
\be\label{Idef2}
I_3(k)=\Delta^{-2} \int_{-\infty}^\infty dx\,x\,\sin\left(\frac{k}{\Delta}\,x\right)\,J_3[\Delta\,Y(x)]\simeq  \frac{\Delta}{48}\int_{-\infty}^\infty dx\,x\,\sin\left(\frac{k}{\Delta}\,x\right)\,Y(x)^3 ,
\ee
where $Y(x)$, in contrast to $Z(y)$, satisfies an equation of motion dependent on the combination $\kappa=\Lambda^{-1}\,\Delta^{-2}$, rather than on the parameters $\Lambda$ and $\Delta$ separately~\cite{ESVW}.
Now clearly, at fixed $\kappa$ and $\Delta\ll1$, the factor $\sin(k\,x\,/\,\Delta)$ undergoes fast oscillations, while $Y(x)$ is independent of $\Delta$, resulting in extensive cancellations. This makes the numerical calculation of $I_3$ difficult, but not impossible. The calculation of integral $I_3$, defined in (\ref{Idef2}), is discussed in Appendix II.

The total transition rate per unit time is
\be\label{square}
\Gamma_3=\int \frac{d^3p}{(2\,\pi)^3E_p}|{\cal M}_3|^2\simeq\frac{1}{8\,\pi^2}\frac{f_a{}^2}{m_a{}^2}\int d^3p\,\frac{1}{E_p}\,\delta(3\,E_0-E_p)\,\frac{1}{k^2}\,[I_3(k)]^2 
\ee
The dimensionless quantity, $k=p\,/\,m_a$, of the ejected axion is  
\be\label{wavenumber}
k_3=\frac{\sqrt{9\,E_0{}^2-m_a^2}}{m_a}=\sqrt{8-9\,\Delta^2}\,\simeq \sqrt{8}.
\ee
Then using the delta function and (\ref{wavenumber}) to integrate over three-momentum in (\ref{square}), we obtain
\be\label{rate}
\Gamma_3=\frac{f_a{}^2}{2\,\pi\,m_a\,k_3}\,[I_3(k_3)]^2
\ee

There are further leading order contributions to the decay rate of axion stars.  Notice that the diagram of Fig. \ref{decaydiagram} can be generalized in such a way that $2\,n+1$ bound axions, where $n=2,\,3,...$, are absorbed and a free axion is ejected.  The generalization of the transition rate to such processes is very simple, 
\be\label{nrate}
\Gamma_{2\,n+1}=\frac{f_a{}^2}{2\,\pi\,m_a\,k_{2\,n+1}}\,[I_{2\,n+1}(k_{2\,n+1})]^2,
\ee
where
\be\label{Indef}
I_{2\,n+1}(k)=\int_{-\infty}^\infty dy\,y\,\sin(k\,y)\,J_{2\,n+1}[Z(y)]
\ee
and
\be\label{nk}
k_{2\,n+1}=\sqrt{4\,n\,(n+1)-[(2\,n+1)\,\Delta]^2}.
\ee
Our calculations show, however, that the contribution of $\Gamma_{2\,n+1}$, for $n\geq2$, to the decay of axion stars is orders of magnitude smaller (and rapidly decreasing with $n$), compared to the leading contribution, $\Gamma_3$. The reason for this is that the rates acquire an additional factor of $\Delta$ for each additional bound axion participating in the process. Therefore, we will not consider those processes any further in the calculation of the lifetime of axion stars.

The only other leading-order, local corrections are the ones in which a number of bound axions are absorbed and two or more free axions are emitted at a single vertex, ${\cal M} =\langle N|V(\Phi)|N-m,\,p_1,...,p_\mu\rangle$, $\mu>1$.   Detailed discussion of the transition rate generated by these processes will be provided in Appendix I.  The most significant difference compared to the transition rate, which we calculated above, is an additional factor of $(m_a{}^2\,/\,f_a{}^2)^{\mu-1}$, which renders the production rate of several free axions in a single process overwhelmingly small. $m_a{}^2\,/\,f_a{}^2\simeq 3\times10^{-52}$ for QCD and it is similarly small in all conceivable physical theories.  Therefore, we disregard these processes for the purpose of calculating the lifetime of axion stars in the following section.

\section{The lifetime of axion stars} \label{lifetime_Sec}
The average time for the  star to shed 3 axions is then $\tau_3=1\,/\,\Gamma_3$.  The axion star contains $N = M \,/ (m_a\,\sqrt{1-\Delta^2})$ axions, where $M$ is the mass of the axion star. Then, using $\Delta\ll 1$, the approximate lifetime of axion stars can be calculated using
\be\label{life}
\frac{d\tau}{dN}\simeq m_a\frac{d\tau}{dM}\simeq-\frac{1}{3}\, \tau_3=-\frac{1}{3\,\Gamma_3}.
\ee

 While, as we will soon see, $1\,/\,|I_3|^2 \ggg 1$ at $\Delta\ll\,1$, at higher values of $\Delta$ the integral $I_3$ approaches a finite limit. Then the integral providing the lifetime of an axion star of mass $M$, obtained from (\ref{life}),\begin{equation}\label{life3}
\tau\simeq\frac{1}{3\,m_a}\int^M dM'\frac{1}{\Gamma_3}=\frac{2\,\pi\,k_3}{3\,f_a{}^2}\int^M dM'\frac{1}{|I_3{}|^2}
\end{equation}
can be cut off at an essentially arbitrary lower limit.

In analyzing the dependence of integral $I_3$, (\ref{Idef2}), on our parameters, consider that the solution of (\ref{eom0}) requires fixing a single integration constant, while the remaining three integration constants are fixed by requiring that the functions $A,\,B,\,\text{and } Z$ are regular at the center of the star. This single nontrivial integration constant is usually chosen as the central density $Z(0)^2$, but can instead be chosen as $\Delta$, or $M$, or  $N$, as these quanities  have a one-to-one relationship with the central density at fixed $f_a$ and $m_a$.  Then $I_3$ only depends on $\Delta$.  However, considering the unique relationship of $M$ and $\Delta$ it can also be considered as a function of $M$, only.

There is another simplification in finding the transition from decaying to surviving axion stars. We find that the transition from stable to unstable axion stars occurs at $\Delta\simeq .05-.06$ in the whole range of $m_a$ we consider.  It is thus easy to see that the transition occurs at $\kappa\ll1$.  Then in (\ref{eomY}) the gravitational field  almost completely decouples from the axion field and we are able to  use  the single equation, (\ref{eomY0}) and, as explained in Appendix II, we find the following analytic expression for the integral $I_3(\sqrt{8})$  at small $\Delta$:
\be\label{I3}
I_3(\sqrt{8})= i\,\frac{32\,\pi\,\rho}{3\,\Delta} \exp\left(-\frac{\sqrt{8}\,\rho}{\Delta}\right).
\ee
The value $\rho\simeq 0.603156$ is obtained by analyzing the Taylor series around $x=0$ and the numerically integrated form of the wave function, $Y(x)$.

Let us consider briefly how one can derive an analytic expression for $\tau$ in (\ref{life3}) in the $\kappa\ll1$ limit.  With boundary conditions $Y(0)=$ finite and $\lim_{x\to\infty} Y(x) =0$, the solution of (\ref{eomY0}) is unique. Then the mass of the star, in the $\Delta\ll1$ limit, is given by~\cite{ESVW}
\be\label{Mdef}
M=\int T_{00} \,d^3r \simeq \frac{\pi\,f_a{}^2}{m_a\,\Delta} \,\int_0^\infty dx\,x^2\,Y(x)^2=y_M\,\frac{f_a{}^2}{m_a\,\Delta},
\ee
where $y_M\simeq25.46$.  Substituting (\ref{Mdef}) and (\ref{I3}) into (\ref{life3}), we obtain $\tau$ as an exponential integral over $M$.  Then introducing the dimensionless variable
\be\label{xidef}
\xi=\frac{2\,\sqrt{8}\,\rho\,m_a\,M}{y_M\,f_a{}^2},
\ee
we obtain the following expression for $\tau$
\be
\tau=\frac{3 \,y_M}{32\,\pi\,\rho\,m_a}\int_{\xi_0}^\xi \frac{d\xi'}{\xi'^2}e^{\xi'},
\ee
where we have introduced an infrared cutoff $\xi_0$.
Now it is easy to see from (\ref{Mdef}) and (\ref{xidef}) that $\xi\gg 1$ when $\Delta\ll1$, so the leading term of the integral is
\be\label{life4}
\tau\simeq\frac{3 \,y_M}{32\,\pi\,\rho\,m_a}\frac{1}{\xi^2}e^{\xi}
      =\frac{3\,y_M{}^3\,f_a{}^4}{1024\,\pi\,\rho^3\,m_a{}^3\,M^2} 
\exp\left(\frac{2\,\sqrt{8}\,\rho\,m_a\,M}{f_a{}^2\,y_M}\right).
\ee 

\begin{figure}[hbt]
\centering\includegraphics[scale=1]{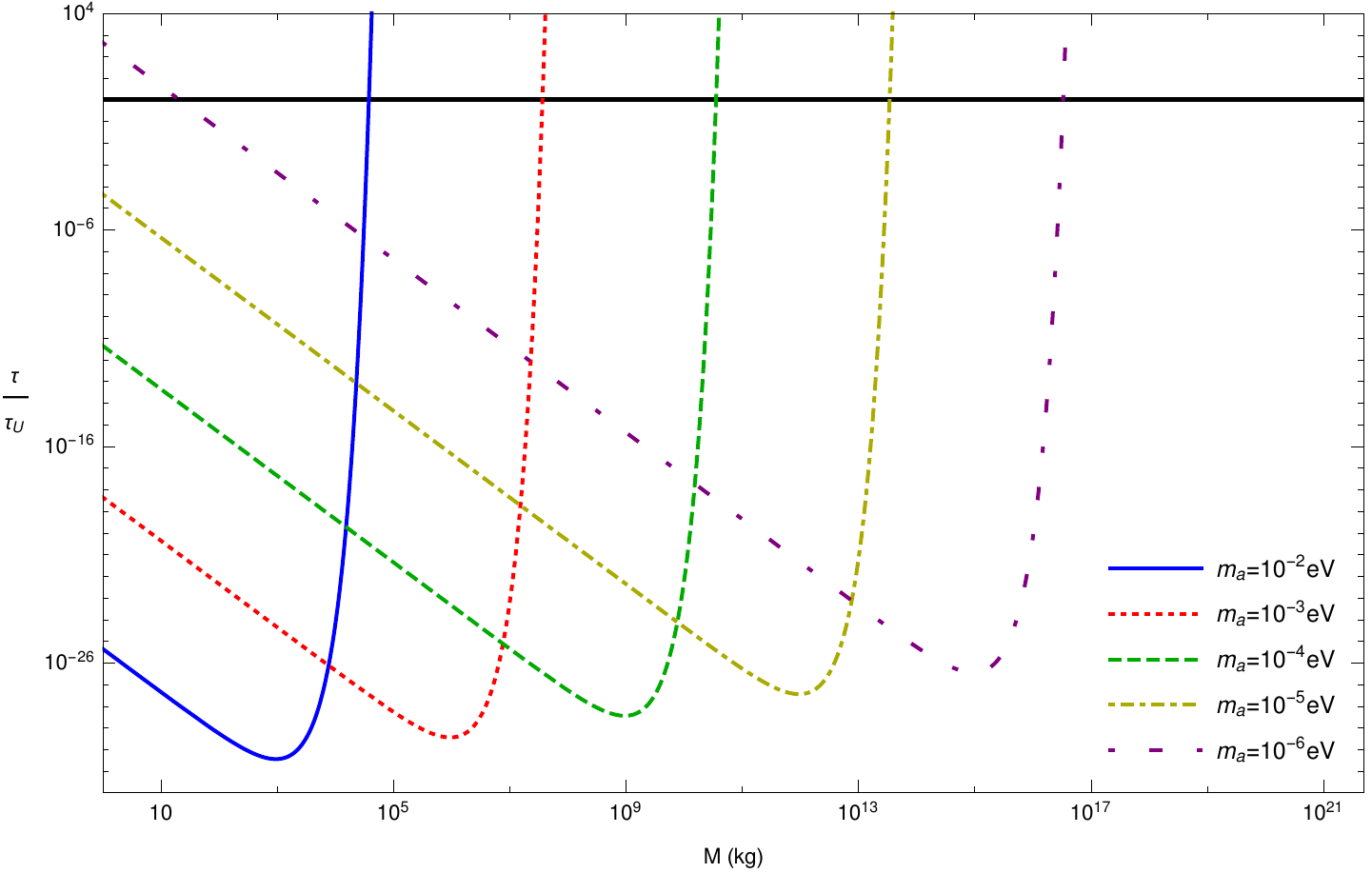}
\caption{The lifetime of axion stars (in units of the age of the universe $\tau_U$) as a function of axion star mass, for different choices of axion particle mass $m_a$.}\label{Lifetimeplots}
\end{figure}

Fixing the QCD relationship  $f_a\,m_a\simeq 6\times 10^{-3} \,\text{GeV}^2$, we equate (\ref{life4}) with the age of the universe ($\tau_U=4.3\times 10^{17}$ s) to find the region of stable masses $M$ as a function of $m_a$.
 In Fig. \ref{Lifetimeplots}, the solid black line represents $\tau = \tau_U$. At large $M$, this sets a lower bound on axion star masses which survive from the big bang to the present day. Because of the exponential dependence of $\tau$ on $M$, even axion stars that survive for only a small fraction of $\tau_U$ (e.g. if they were produced at relatively late times) have a similar lower bound on their masses. At small masses, as depicted in the top-left corner of Fig. \ref{Lifetimeplots}, $\tau > \tau_U$ sets an upper bound which is $\mathcal{O}(10)$ kg or less (depending on $m_a$), so that very light axion stars, while less interesting phenomenologically, could be stable as well. In the analysis, we have concentrated on the mass range $10^{-6} \text{ eV}\leq m_a\leq 10^{-2} \text{ eV}$, most of which is not ruled out by observations~\cite{observe1,observe2}.
 
\begin{figure}[hbt]
\centering\includegraphics[scale=.7]{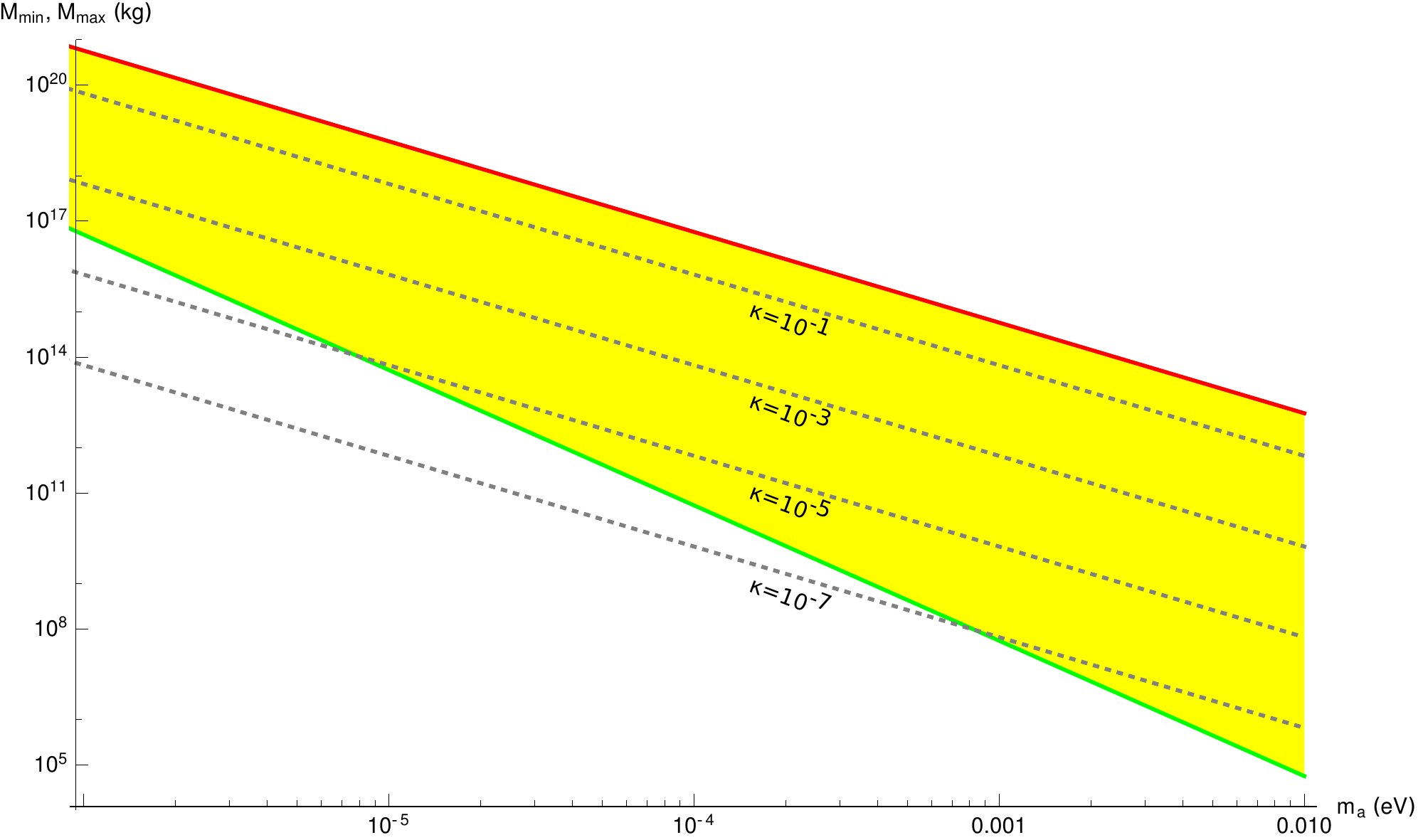}
\caption{Upper (gravitational stability, red line) and lower (stability against axion emission, green line) bounds for the masses of axion stars, as  functions of the axion mass in QCD. The dotted lines correspond to different values of the $\kappa$, which parametrizes the effective coupling to gravity.}\label{QCDplots}
\end{figure}

In Fig. \ref{QCDplots} we plotted the limits of gravitational stability (red line) and stability against axion decay (green line) as functions of $m_a$ and $M$ in QCD.  Gravitational stability of bosonic condensates was studied in~\cite{Gleiser}. A conclusion of that study was that regions of gravitational stability of objects similar to axion stars end at  maximum mass configurations. We arrived at similar conclusion by studying the binding energy as a function of mass~\cite{ESVW}.  Axion stars having mass in the yellow area are stable against both of these instabilities.  Axion stars have maximal masses for a given $m_a$ along the red line, and have effective coupling to the gravitational field, $\kappa\simeq0.34$.  Axion stars along the green line are almost completely decoupled from gravity, having $\kappa\ll1$. Such axion stars are thus held together not by gravitational attraction, but through their attractive self-interactions alone.

Note that once one fixes the product $m_a\,f_a$, the maximal mass scales with $m_a{}^{-2}$, while the minimal mass scales approximately with  $m_a{}^{-3}$. This is easy to see in (\ref{Mdef}): the maximum mass is a curve of constant $\kappa\simeq .34$, while the minimum is a curve of (approximately) constant $\Delta \simeq .05-.06$.

A recent study of microlensing by MACHOS~\cite{Griest} found that masses in the range of $4\times 10^{21} \text{ kg}\leq M\leq 2\times 10^{23}\text{ kg}$ cannot make up the entirety of dark matter in the Milky Way.  However, we find that all stable QCD axion stars have masses $M \lesssim 10^{21}$, safely below these bounds.

Clearly, if axion stars in the stable mass region of Fig. \ref{QCDplots} were produced, they would have emitted a large number of free axions since the time they were created in the early universe.  Further, any smaller axion stars, with masses below the green line in Fig. \ref{QCDplots}, would have decayed completely and also produced a large number of free axions.  The current mass distribution of axion stars is a function of the primordial axion star mass distribution. 

The relativistic axions produced by the decay process escape from the galaxy and galaxy clusters. This means that if axion stars provide a substantial fraction of dark matter, then the relative amount of dark matter contained in those structures would decrease in time. If such a decrease can be detected by observations, this would place limits on the abundance of axion stars. 

Observational bounds on the axion mass include studies of white dwarf cooling~\cite{dwarf1,dwarf2} and supernova burst durations~\cite{Raffelt}, which result in excluding axions of mass $m_a\geq 1.6 \times 10^{-2}\text{ eV}$. However, these numbers are somewhat model dependent. Lower limits on the axion mass, $m_a \geq 10^{-5}\text{ eV}$~\cite{fourth}, are provided mostly assuming that free axions constitute the entirety of cold dark matter.

 \section{Summary}
 One may na\"ively assume that  axion stars, which are composed of self-adjoint bosons, may not exist in nature because the axion number is not conserved; consequently, one would conclude that axion stars cannot contribute to dark matter. Alternatively, one may tacitly assume that decay through the self-interaction of axions is negligible.  We have shown that the truth is between the above extremes and obtained interesting bounds on the mass of surviving axion stars. Our method can be applied to any boson star that is formed from condensed real bosons.
 
 Though axions decay into photons, it has been shown to have an insignificant effect on axions stars  during the existence of the universe~\cite{Tkachev-1}.  
  In this paper, we have estimated the lifetime of axion stars which decay through the self interaction of axions. We have done that using lowest order perturbation theory, which is an excellent approximation, due to the smallness of the $2\,n$ axion coupling constant, $\lambda_n\sim \left(m_a{}^2\,/\,f_a{}^2\right)^{2\,n-2}$.  Even $\lambda_2$ is very small, $\mathcal{O}(10^{-50 })$ for QCD axions, and is  small for GUT or string axions as well. 
  
 Then the most important decay process is $3\,a\to a$, where the initial state axions are bound in their ground state, and the final state axion is free. While energetically clearly admissible, this process is possible only because the bound axions are not in momentum eigenstates.  The size of their wave packet is roughly the inverse of the radius of the axion star,  $R\sim 1\,/\,(m_a\,\Delta)$.  At strong binding $\Delta =\mathcal{O}(1)$, the uncertainty of the momentum is $\delta p=\mathcal{O}(m_a)$ and three bound axions easily produce $p\sim m_a$, the momentum carried away by the free axion, as required by energy conservation.   This results in a large transition rate, $\Gamma\sim m_a$ and a very short lifetime for axion stars.  
 
 In conclusion, small axion stars produced after the big bang rapidly decay and do not survive until the present time. However, the heaviest axion stars (which are still gravitationally stable) are weakly bound and their radius is very large compared to the Compton wavelength $1/m_a$.  The uncertainty of the momentum of the bound axions is reduced to $\delta p\sim m_a\, \Delta$, making production of a free axion increasingly improbable. Since the tail of the momentum distribution $f(p) \sim \exp\left(- p\,R\right)$ cuts off exponentially (as explained in Appendix II), the transition rate decreases rapidly with decreasing $\Delta$. At $\Delta \simeq 0.05 - 0.06$ it becomes sufficiently small such that axion stars survive until the present time. 
  
Axion stars become metastable if the effective coupling of axions to gravitation, $\kappa = ( f_a\,/M_P\,\Delta)^2\gtrsim 0.34$ ~\cite{ESVW} (see also ~\cite{Gleiser}).  For $\kappa\lesssim 0.34$ the transition rate can still be small enough that some stars produced in the early universe would survive until the present time. 

We pointed out in the previous section that if a substantial fraction of dark matter is composed of weakly bound axion stars, then the ratio of dark matter to luminous matter becomes a function of time.  (Note that axions emitted from axion stars are relativistic and they escape from galaxies and from galaxy clusters.)  Thus, by comparing the rotation curves of galaxies of different ages one should be able to put an upper limit on the contribution of weakly bound axion stars to dark matter.
In a future publication we will also consider condensates of axions not originating in QCD, inspired by string theory, which may be as large as  galaxies, or even as the whole universe~\cite{lightaxion1,lightaxion2}.
\\

{\bf Acknowledgements}\\
We thank P. Argyres, J. Brod, M. Ma, C. Prescod-Weinstein, T. Takeuchi, and C. Vaz for conversations. J.E. is supported by a Mary J. Hanna Fellowship.  L.C.R.W. acknowledges support by the Aspen Center for Physics through the NSF Grant PHY-1066293.

 \renewcommand{\theequation}{A-\arabic{equation}}
   \setcounter{equation}{0}
  \section*{Appendix I. $\mu$ axion emission.}  
We calculated the matrix element   
\be\label{generaln}
\left\langle N \left\vert 1- \cos\left(\frac{\Phi}{f_a}\right)\right\vert N-n \right\rangle\simeq\left\langle N \left\vert e^{i\,\Phi^+\,/\,f_a}e^{i\,\Phi^-\,/\,f_a}\right\vert N-n \right\rangle=-i^n\,J_n\left[Z(y)\right] e^{i\,n\,E_0}
\ee
in~\cite{ESVW}, in leading order of $N$. The states $\langle N \vert$ and $\vert N-n\rangle$ are ground state condensates of $N$ and $N-n$ axions, and $\Phi^{+,-}$ are the first term of (\ref{field2}) and its Hermitian conjugate.   Since the first two terms of (\ref{field2}) commute, we also obtain
\be \label{expect}
\left\langle N \left\vert 1- \cos\left(\frac{\Phi}{f_a}\right)\right\vert N-n,\,p_1,...,p_\mu \right\rangle
=-i^n\,J_n\left[Z(y)\right]\,e^{i\,n\,E_0}\left(\frac{i}{f_a}\right)^\mu\,\prod_{s=1}^{\mu}\frac{e^{i(\vec p_s\cdot \vec r - E_s\,t) }}{\sqrt{2\,E_s}},
\ee
where the state $\vert N-n,\,p_1,...,p_\mu \rangle$ contains a star of $N-3$ condensed axions plus $\mu$ free axions of momentum $p_s$ and energy $E_s$. Multiplying (\ref{expect}) by the scale factor $m_a{}^2\,f_a{}^2$ and integrating over $\vec{r}$ and $t$ gives the matrix element
\be\label{matrixgeneral}
{\cal M}=-2\,\pi\,m_a{}^2\,i^n\left(\frac{i}{f_a}\right)^{\mu-2}\,\delta(n\,E_0-\sum_s E_s)\,\int d^3r\,J_n(Z)\exp\left(i\,\vec r \cdot\left[\sum_s\vec p_s\right]\right)\,\prod_{s=1}^\mu\,\frac{1}{\sqrt{2\,E_s}}.
\ee

The application of (\ref{matrixgeneral}) to the simplest case of $n=3$ and $\mu=1$ has been discussed in Sec. 3, which gave (\ref{matrix2}). Here we will discuss the emission of two axions, $\mu=2$.  As shown by (\ref{matrixgeneral}), 
$|\mathcal{M}|^2$ is proportional to $f_a{}^{4-2\mu}$.  Since the remainder of the transition rate scales only with $m_a$, the rate of emission of $\mu$ axions compared to that of a single axion is $r_\mu=\Gamma^\mu\,/\,\Gamma^1\sim (m_a\,/\,f_a)^{2\mu-2}$. This is a minuscule factor for almost every conceivable physical theory, e.g. for QCD, $m_a{}^2\,/\,f_a{}^2\sim 10^{-50}.$ We sketch the calculation of the rest of the matrix element for $\mu=2$, to show that it does not come close to compensating the smallness of $r_\mu$.

The most probable process for the emission of two free axions is the transformation of four bound axions into two free axions. In that case, the matrix element takes the form
\be
{\cal M}=\frac{4\,\pi^2}{m_a\,K\,\sqrt{E_1\,E_2}}\,I_4(K)\,\delta(4\,E_0-E_1-E_2),
\ee
where $K=|\vec p_1+\vec p_2|\,/\,m_a$ and where $I_4(K)$ was defined in (\ref{Indef}).

Then the transition rate per unit time is
\be\label{rate41}
\Gamma_4^2=\frac{1}{2\,\pi^4\,m_a{}^2}\int\frac{d^3p_1}{E_1}\,\frac{d^3p_2}{E_2} \frac{1}{K^2}[I_4(K)]^2\,\delta(4\,E_0-E_1-E_2),
\ee
where $\Gamma^\mu_n$ is the transition rate of the process  $N \,a_0\to (N-n)\,a_0+ a_{k_1}+...+a_{k_\mu}.$
The integration over every variable but $K$ can be performed analytically, to give
\be\label{rate42}
\Gamma_4^2=\frac{m_a}{16\pi^3} \int_0^{\sqrt{16\,\epsilon_0{}^2-4}}dK\, f(K)\,[I_4(K)]^2,
\ee
where $f(K)$ is a complicated function involving logarithms of expression containing $K$. However, at weak binding, $\epsilon_0\to 1$,  $f(K)$ simplifies to
\be
f(K)\simeq \frac{K\,(12-K^2)}{16-K^2},
\ee
while the upper limit of integration over $K$ becomes $\sqrt{12}$. 

We calculated the the ratio of $\Gamma_4^2\,/\Gamma_3^1$, where $\Gamma_3^1$ is given in (\ref{rate}), at several choices of the input parameters and found that 
\be
\frac{\Gamma_4^2}{\Gamma_3^1} = C \,\frac{m_a{}^2}{f_a{}^2},
\ee
where $C$ is a constant of $\mathcal{O}(1)-\mathcal{O}(10)$.  The  ratio $m_a{}^2\,/\,f_a{}^2 \ll 1$ makes the contribution of the process of emission of more than 1 axion to the decay of axion stars negligible.
  \section*{Appendix II. The integral $I_3$}  

Regular  solutions of (\ref{eom0}) are also quite likely entire functions. Solutions of the nonlinear system (\ref{eomY}) that are regular at $x=0$ and vanish at infinity are uniquely defined and are real analytic, but have singularities at pairs of  complex conjugate points.  To find the nature of these singularities, we expand the equation of motion (\ref{eomY0}) around the singular term, taking the ansatz
\be\label{Ysing}
Y_{\rm sing}\simeq \frac{A}{(x^2+\rho^2)^\alpha}.
\ee
The only nontrivial solution is found if  $\alpha=1$. Then the singularities are first order poles and in leading order of the Laurent series expansion around $x^2=-\rho^2$ we obtain the following constraint on parameters $A$ and $\rho$ : $A=\pm\,8\,\rho$.  

The nature and location of the singularities with the smallest distance from the real axis (denoted $\rho$) can also be found out from examining the Taylor series expansion of the solution $Y(x)$ around $x=0$
\be
Y(x)=   \sum_{n=0}^N \eta(n) \,x^{2\,n}.
\ee
Assuming a leading singularity of the form (\ref{Ysing}), it is easy to deduce the value of $\alpha$, $A$ and $\rho$, using the asymptotic expansions in $n$, 
\bea
\alpha&=&1+n^2\left(1-\frac{\eta(n+1)\,\eta(n-1)}{\eta(n)^2}\right)+O(n^{-4}),\label{alpha}\\
\rho&=&\sqrt{-\frac{\eta(n-1)}{\eta(n)}}\left(1+\frac{1-\alpha}{n}+\mathcal{O}(n^{-2})\right),\label{rho}\\
A&=&(-1)^n\,\eta(n)\,\rho^{2\,n+2}\,\alpha\, \frac{\alpha+1}{2}\,\frac{\alpha+2}{3}\left(1+\frac{\alpha-1}{n}\right).\label{A}
\eea

The expression (\ref{alpha}), evaluated at $n=20,\,19,\,\text{and}\,18$, gives $\alpha-1=2.1\times 10^{-6},\,2.6\times 10^{-6},\,\text{and}\,3.2\times 10^{-6}$, respectively, implying that $\alpha=1$ (i.e. the singularity is a single pole) with very high confidence.  Then (\ref{A}) reduces to $A=(-1)^n\,\eta(n)\,\rho^{2\,n+2}$ and gives $A=4.81747$, while (\ref{rho}) gives $\rho=0.602184$, for each of $n=20,\,19,\,\text{and}\,18.$  So the relationship $A=8 \,\rho$ is also precisely satisfied.  Thus, we have shown that the closest singularity of $Y(x)$ is of the form  (\ref{Ysing}) with $\alpha=1$, and the distance of the pole from the real axis is $\rho=0.602184$. 

Consequently, the contour of integration in $I_3$, which can be cast in the form
\be
I_3=\frac{\Delta}{48}\int_{-\infty}^\infty dx\,x\,\exp\left(i\,\frac{k}{\Delta}\,x\right)\,Y(x)^3 ,
\ee
can be deformed until we encounter the first singularity, at $i\,\rho$, in the upper half complex plane. The contribution of the first pole in the upper half plane dominates the integral which, using (\ref{Ysing}), evaluates to the following expression in leading order of $\Delta$:
\be\label{I3final}
 I_3\simeq \frac{32\,i\,\pi\,\rho}{3\,\Delta}\,\exp\left(-\frac{\sqrt{8}\,\rho}{\Delta}\right).
 \ee
 
 We also calculated $I_3$ by brute force, using the wave function obtained from numerical integration of (\ref{eomY0}). This can only be done for moderately small values of $\Delta$. The expression (\ref{I3final}) fits those calculations with the choice $\rho\simeq0.6$ extremely well. 

\setcounter{equation}{0}

\end{document}